\documentclass[aps,pre,letterpaper,reprint]{revtex4-2}
\usepackage{latexsym,eucal,amsfonts,amsmath,amsthm,amssymb,nccmath} 
\usepackage{scalerel}
\usepackage{blindtext}
\usepackage[table]{xcolor}
\usepackage{graphicx}
\usepackage{psfrag}
\usepackage{amsbsy}      
\usepackage[switch]{lineno}
\usepackage[draft]{changes}

\usepackage{comment}
\usepackage{siunitx}
\usepackage{chemfig}
\usepackage{mathtools}
\usepackage{xfrac}
\usepackage{color}
\usepackage{dsfont}
\usepackage{tikz}
\usepackage{bm}
\usepackage[utf8]{inputenc}
\usepackage[T1]{fontenc}

\def\c{{\boldsymbol c}}
\def\dd{\mbox{d}}

\def\x{{\boldsymbol x}}

\def\n{{\bf n}}
\def\r{\boldsymbol r}

\def\z{{\boldsymbol z}}

\def\0{{\bf 0}}

\def\W{{\bf W}}

\begin{document}
\title{Incorporating stochastic gene expression, signaling-mediated
  intercellular interactions, and regulated cell proliferation in
  models of coordinated tissue development}

\author{Casey O. Barkan}
\email{barkanc@ucla.edu}
\affiliation{Department of Physics \& Astronomy, 
University of California, Los Angeles, CA 90095-1766 USA}
\author{Tom Chou}
\email{tomchou@ucla.edu}
\affiliation{Department of Mathematics, University of California, Los Angeles, CA
90095-1555 USA}


\begin{abstract}

Formulating quantitative and predictive models for tissue development
requires consideration of the complex, stochastic gene expression
dynamics, its regulation via cell-to-cell interactions, and cell
proliferation. Including all of these processes into a practical
mathematical framework requires complex expressions that are difficult
to interpret and apply. We construct a simple theory that incorporates
intracellular stochastic gene expression dynamics, signaling chemicals
that influence these dynamics and mediate cell-cell interactions, and
cell proliferation and its accompanying differentiation. Cellular
states (genetic and epigenetic) are described by a Waddington vector
field that allows for non-gradient dynamics (cycles, entropy
production, loss of detailed balance) which is precluded in Waddington
potential landscape representations of gene expression dynamics. We
define an epigenetic fitness landscape that describes the
proliferation of different cell types, and elucidate how this fitness
landscape is related to Waddington's vector field. We illustrate the
applicability of our framework by analyzing two model systems: an
interacting two-gene differentiation process and a spatiotemporal
organism model inspired by planaria.
    
\end{abstract}

%

\maketitle

\section{Introduction}\label{SEC:INTRO}

Cell and tissue development is a complex and multifaceted dynamical
process, involving gene regulatory dynamics, cell-to-cell
interactions, and differential cell proliferation and death rates. The
conceptual model of Waddington's landscape has been a guiding paradigm
for understanding the mechanisms of cell development
\cite{waddington1942canalization,ferrell2012bistability,schiebinger2021reconstructing}. Waddington's
landscape suggests that the internal dynamics of a differentiating
cell follow that of a particle in a rugged high-dimensional
``potential'' landscape.  However, the internal dynamics of cell
states (\textit{e.g.}, gene expression profiles) are only one aspect
of tissue development.  Proliferation and death of individual cells
give rise to population dynamics akin to that seen under Darwinian
evolution \cite{shahriyari2013symmetric,lei2020general}. Additionally,
intracell gene dynamics and population-level birth-death processes are
coupled through cell-to-cell signaling and metabolic interactions
\cite{lei2020general,lei2023mathematical}. Thus, a complete
description of development must describe an interacting population of
cells, not just the gene dynamics within a single cell.

The conceptual framework of Waddington's landscape has been translated
into a variety of mathematical models.  For example, Waddington's
landscape has been interpreted as a potential landscape acting on
single-cell gene expression
\cite{wang2008potential,schiebinger2021reconstructing,rand2021geometry,coomer2022noise,saez2022statistically,zhou2016construction,lei2023mathematical}
and on population level phenotypes \cite{raju2023theoretical}. In the
typical single-cell interpretation, a cell's state is modeled as a
particle with position $\x$ in a potential energy landscape $U(\x)$ in
a noisy environment. This model implies that one can reconstruct
$U(\x)$ from data using the simple relationship $U(\x)=-\log p_{\rm
  ss}(\x)$, where $p_{\rm ss}(\x)$ is the steady-state distribution of
states empirically measured.  Such formulae have been applied to real
datasets \cite{sisan2012predicting} but its use is hampered by
its inherent limitation to a single cell (or to non-interacting and
non-proliferating populations) and the assumption of both gradient
dynamics \textit{and} dynamics that obey detailed balance among
gene regulation states.


When cell division and death are considered, the concept of a fitness
landscape becomes relevant. In evolutionary population dynamics, high
fitness genotypes elicit high proliferation rates; this concept must
be adapted to the context of development, where all cells have the
same genotype. Mathematical models of development often consider a
single-cell fates and state changes
\cite{wang2011quantifying,sisan2012predicting,fard2016not,
  rand2021geometry,schiebinger2021reconstructing,qiu2022mapping,
  frank2023macrophage,nakamura2024evolution,xue2024logic,
  chen2025quantifying,XING_ELIFE,XING_PRXL}; those that incorporate
cell population dynamics through birth and death typically neglect
cell-cell interactions
\cite{weinreb2018fundamental,fischer2019inferring,
  shi2019quantifying,zhang2021optimal,forrow2021lineageot,
  shi2022energy}. In kinetic theories that combine internal cell state
dynamics with state-dependent division and death rates, tractable
dynamical equations describing population-level quantities can be
derived if there are no cell-cell interactions
\cite{Xia_JPA,Xia_PRE,Kinetic_PRE,Kinetic_JSP}.  However, in
developing tissue, cell division and death are tightly coordinated
through cell-cell interactions mediated by secreted chemical signals
that influence gene expression in other cells
\cite{GILBERT,discher2009growth}. Failure of this regulated tissue
development can lead to pathologies such as cancer
\cite{huang2013genetic,huang2021reconciling}. Thus, effective
theoretical approaches that incorporate regulated intracell gene
expression with population-level birth-death processes will provide
critical tools for modeling and analyzing tissue development and
evolution.

In this paper, we present a general framework for modeling a general
developmental process in which cell-cell interactions regulate gene
expression and cell proliferation.  Our framework describes a
population of individual (discrete) cells, rather than assuming an
infinite continuum of cells, as has been done in other works
\cite{zhang2021optimal,shi2019quantifying,lei2020general,lei2023mathematical}.
In the tissue development context, our framework also includes spatial
structure, enabling models of coordinated tissue development in which
a small initial population of stem cells robustly develops into a
stable and self-healing tissue. Our approach incorporates all known
important processes in an interpretable way, allowing us to better
define the concepts of the Waddington landscape and fitness landscape
in the context of development. Rather than use a Waddington landscape
that assumes gradient dynamics, we use a \textit{Waddington vector
  field} to account for non-gradient dynamics of gene expression. Such
detailed-balance-violating stochastic dynamics is expected in an
energy-consuming process such as cell state regulation and
homeostasis.

We also define an effective \textit{epigenetic fitness
  landscape}. Whereas a fitness landscape describes differential
proliferation of different genotypes, our epigenetic fitness landscape
describes the differential proliferation of cells of the same genotype
type but differing epigenetic state, capturing the fact that different
attractors of the gene regulatory dynamics induce different cell
division and death rates. In addition, we show how our general model
can be reduced to simpler forms by applying common assumptions. Often,
such assumptions are implicitly included in models of development,
leading to confusion regarding the realm of applicability of such
models. Our general framework provides a starting point for deriving
simpler, interpretable models by applying explicitly-stated
assumptions.  We present two models of the robust development of a
stable tissue population as proof-of-concept examples. The first is a
well-mixed model of gene-regulated stem cell differentiation, and the
second is a spatiotemporal model of a regenerating organism, similar
to planaria.

\section{Mathematical Framework}

Here, we assemble all the relevant physical processes that are
understood to play roles in a developing tissue. The dependencies (and
feedback) across scales and among these mechanisms are clearly
delineated by proper definitions and judicious
approximations/assumptions.

\subsection{Intracellular state dynamics}

Let $\z\in\mathbb{R}^{d_z}$ denote the internal state of a cell. $\z$
specifies the state of the relevant gene regulatory network(s) in the
cell, and, in principle, could also specify the methylation of DNA,
chromatin structure, spatial organization of organelles, and/or any
other relevant variables. While a complete specification of the gene
regulatory network state requires specification of both protein and
mRNA concentrations within the cell, the relatively fast timescales of
mRNA transcription and degradation compared to protein translation and
degradation can justify a quasi-steady state approximation where only
protein concentrations need be specified. We label cells by a
superscript $\alpha$, \textit{i.e.}, $\z^\alpha$ is the state of cell
$\alpha$, where $\alpha=1,...,N(t)$, and $N(t)$ is the total number of
cells which varies from cell division and death \footnote{The labeling
of cells is arbitrary. One straightforward labeling scheme is to label
cells from oldest to youngest, \textit{i.e.}, cell $\alpha=1$ is the
oldest cell. When a cell dies, all cells younger than the dying cell
are then re-labeled.}.
%
%

To allow for spatial resolution, let $\r^\alpha\in \mathbb{R}^3$
denote cell $\alpha$'s position in 3-dimensional space. Interactions
between cells are typically mediated by signaling molecules; suppose
there are $L$ relevant signaling molecules and let $c_{\ell}(\r,t)$
denote the concentration of signaling molecule $\ell$ ($\ell=1,...,L$)
at position $\r$ at time $t$. For notational convenience, let
$\c(\r,t)=(c_1(\r,t) ,...,c_{L}(\r,t))$ denote the list of signaling
molecule concentrations. For cell $\alpha$ at position $\r^\alpha$,
the dynamics of its state $\z^\alpha(t)$ are modeled by a stochastic
differential equation (SDE):
\begin{equation}\label{x_dynamics}
  \dd \z^\alpha = \bm{F}(\z^\alpha; \c^\alpha)\dd t
  + \sigma(\z^\alpha; \c^\alpha) \dd \bm{B}^\alpha,
\end{equation}
where $\c^\alpha\equiv \c(\r^\alpha,t)$ are the signaling concentrations
experienced by cell $\alpha$ at time $t$. $\bm{F}(\z;\c)$ defines a
$\c$-dependent vector field that maps
$\mathbb{R}^{d_z}\to\mathbb{R}^{d_z}$ and which describes the
deterministic component of the cell state dynamics. We call
$\bm{F}(\z; \c)$ \textit{Waddington's vector field}, and it is presumed
to have stable attractors corresponding to stable cell
types. Importantly, the signaling molecule concentration $\c^\alpha$
influences the dynamics of the state $\z^\alpha$, and the dynamics of
$\c(\r,t)$ are governed by the states and positions of all cells (as
described below). Hence, the signaling molecules can mediate
cell-to-cell interactions, allowing the tissue to develop in a
coordinated way. In the special case that $\bm{F}(\z;\c)$ is the
negative gradient of a function, $\bm{F}(\z;\c)=-\nabla_\z U(\z;\c)$,
then $U(\z;\c)$ would be the traditional \textit{Waddington
  landscape}. However, we do not assume that $\bm F(\z;\c)$ is the
gradient of any function, and in section~\ref{landscapes} we review
and discuss the limitations of the traditional Waddington landscape
picture.
%
%
Noise is characterized by the $d_b$-dimensional Brownian noise $\dd
\bm{B}^\alpha$ and by the amplitude $\sigma(\z; \c)$, a $d_{z}\times
d_{b}$ matrix
%
%
which may depend on the cell state and signaling molecule
concentrations that the cell experiences.



\subsection{Cell-cell interactions through signaling molecules}

\begin{figure}
\centering
\includegraphics[width=3.3in]{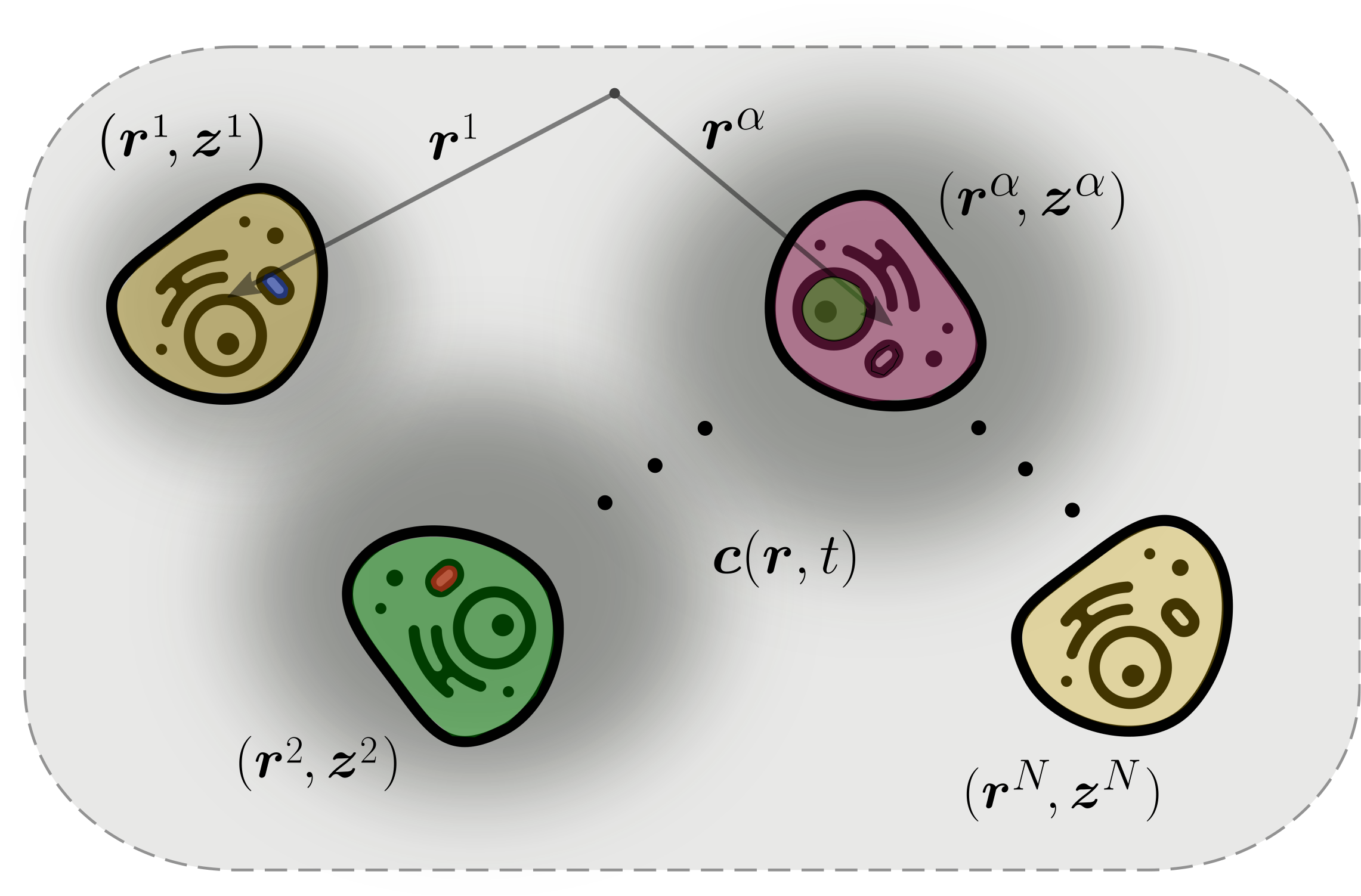}
\caption{Schematic of cells at various positions $\r^{\alpha}$ and in
  various gene expression states $\z^\alpha, \,
  \alpha=1,2,...,N$. Cell-cell interactions mediated by
  position-dependent signaling molecule concentrations $\c(\r,t)
  \equiv \big(c_1(\r,t),...,c_L(\r,t)\big)$ which are represented in
  gray-scale. Cells can produce signaling molecules at rates that
  depend on $\z$; gene expression is in turn influenced by the
  concentrations that a cell senses, leading to potentially complex
  cell-cell interactions.}
\label{signaling_interaction}
\end{figure}

The dynamics of signaling molecule concentrations can be modeled by
any variety of molecular transport models. For example, a
deterministic PDE describing excretion, absorption, and diffusion of
signaling molecules takes on the typical form \cite{fooladi2019enhanced}

\begin{equation}\label{lambda_dynamics}
  \partial_t c_{\ell}(\r,t) = D_{\ell}\nabla^2
  c_{\ell} - d_{\ell}c_{\ell} + \sum_{\text{cells }\alpha}\!\!\lambda_{\ell}(\r-\r^\alpha\!, \z^\alpha),
\end{equation}
where $\lambda_{\ell}(\r-\r^\alpha\!,\z^\alpha)$ is the excretion rate
of signaling molecule $\ell$ at position $\r$ from a cell in state
$\z^{\alpha}$ centered at $\r^{\alpha}$, and the sum runs over all
cells. The function $\lambda_{\ell}(\r-\r^\alpha\!,\z^\alpha)$ is
localized about $\r^{\alpha}$ and represents the spatial extent of a
cell that exports signaling molecule.  Diffusion and degradation of
molecule $\ell$ is described by the coefficients $D_{\ell}$ and
$d_{\ell}$, respectively.  A natural choice of boundary condition for
Eq.~\ref{lambda_dynamics} is $c_{\ell}(\r) \to 0$ as $|\r| \to
\infty$. Alternatively, to model in subcompartments or spatially
defined microenvironments, Eq.~\ref{lambda_dynamics} can be considered
in finite regions of space each with appropriate boundary conditions.
For a closed environment, one would use reflecting boundary conditions
for $c_{\ell}(\r)$.  Eq.~\ref{lambda_dynamics} neglects reactions
amongst the signaling molecules and models deterministic or locally
averaged concentrations, appropriate for molecular transport at high
concentrations and/or fast molecular timescales.  At very low
signaling molecule densities, intrinsic molecular stochasticity may be
important and any number of stochastic models
\cite{KULKARNI,GRIMA,CHEN_JIA} and simulation approaches
\cite{gillespie1977exact,gillespie2000chemical} are available.
Fig.~\ref{signaling_interaction} provides a schematic of
signaling-molecule-mediated interactions among cells at positions
$\r^{\alpha}$, with potentially different internal states
$\z^{\alpha}$.

\subsection{Cell migration}

Cells may also migrate by exerting forces on neighboring cells or on
their substrate; this migration may be influenced by signaling
molecules (chemokines). Such coordinated movement is essential for
tissue development \cite{xi2019material,ranft2010fluidization}. One
can straightforwardly couple molecular signaling with cell motion
\cite{PB_CHOU}, and a general stochastic model for cell motion can be
expressed using an SDE such as
\begin{equation}
  \label{movement}
  \dd \r^\alpha(t) = \bm{g}(\r^\alpha,S(t)) \dd t +
  \eta(\r^\alpha,S(t)) \dd \bm{W}^\alpha
\end{equation}
where $S(t) = \{(\r^\alpha(t), \z^\alpha(t))\}_{\alpha=1}^{N(t)}$ is
the full state of the tissue population and $\dd \bm{W}^\alpha$ is a
3-dimensional Brownian noise. The mean velocity $\bm{g}$ and noise
amplitude $\eta$ (a $3\times3$ matrix) can depend on the cell's
position $\r^\alpha$ as well as on the full state of the tissue
$S(t)$. Although explicit dependence of $\bm{g}$ and $\eta$ on the
spatial profile of signaling molecule concentrations is not included
in Eq.~\ref{movement}, one could consider a velocity $\bm{g}$ which
depends on $\nabla \c(\r^\alpha, t)$, or on higher derivatives of the
concentration profile.  Eq.~\ref{movement} is a Langevin equation for
fluctuating cell positions. Many related models for cell motion,
including anomalous diffusion and longer-ranged hopping have been
developed
\cite{dieterich2008anomalous,harris2012generalized,HOFLING_2013}.

\subsection{Cell proliferation (and death)}

During cell division---a process involving cellular-scale disruption
and reorganization---a mother cell divides into two daughter cells which may
acquire states that are different from the state of the mother. In our
model, cells have a state-dependent and concentration-dependent
division rate, denoted $\beta(\z;\c)$ for a cell in state $\z$ and with
signaling molecule concentrations $\c$ at its location (to be precise,
cell $\alpha$ has division rate $\beta(\z^\alpha;\c^\alpha)$, but we
will drop the superscript $\alpha$ to avoid cumbersome
notation). $\beta(\z; \c)$ is a marginal probability over the joint
distribution for the probability of a cell in state $\z$ dividing into
daughters with states $\z'$, $\z''$ and with positional displacements
from the mother $\Delta \r'$, $\Delta\r''$. This joint distribution is
denoted as

\begin{equation}
  \label{birth}
  \tilde{\beta}\big(\z ; \c ; \z',\z'',
  \Delta \r',\Delta \r''\big)
\end{equation}
and is invariant under exchange of $\z'$ and $\z''$ and
under exchange of $\Delta\r'$ and $\Delta\r''$. Marginalizing, we find the overall division rate
\begin{equation}
 \beta(\z;\c)= \int\!\tilde{\beta}\big(\z ; \c ; \z',\z''\!,
  \Delta \r',\Delta \r''\big)
  \dd \z' \dd\z''\dd\Delta\r' \dd\Delta\r''.
\end{equation}
Finally, cell death occurs at rate $\mu(\z; \c)$, 
%
which can depend on both $\z$ and $\c$.

Eqs.~\ref{x_dynamics}, \ref{lambda_dynamics}, and \ref{movement},
along with system-specific models for birth and death rates,
$\tilde{\beta}(\z ; \c ; \z',\z'', \Delta \r',\Delta \r'')$ and
$\mu(\z; \c)$, form our general description of developing cell
populations. The following section illustrates how common assumptions
can be incorporated to significantly simplify the model.

\section{Simplifying assumptions}\label{assumptions}

Here we describe a series of assumptions that can be applied to
further simplify our mathematical model. Our examples in
Sections~\ref{ex1} and \ref{ex2} show how various combinations of
these assumption can be used to obtain models that capture different
degrees of detail.

\subsection{Fast equilibration of signaling molecule concentrations}\label{fast_c}
Suppose the concentrations $c_\ell(\r,t)$ reach steady state on
timescales shorter than those associated with gene expression
dynamics, cell movement, and cell division and death. We can
approximate $c_{\ell}(\r,t)$ by the quasi-steady state solution of
Eq.~\ref{lambda_dynamics} under the instantaneous state of the
population $S(t) = \{\z^\alpha(t),\r^\alpha(t)\}_{\alpha=1}^{N(t)}$.
We denote this quasi-steady state concentration profile by $c^{\rm
  ss}_{\ell}(\r; S(t))$. For signaling molecule dynamics described by
Eq.~\ref{lambda_dynamics}, $c^{\rm ss}_{\ell}(\r; S(t))$ takes the
explicit form
\begin{equation}\label{c_ss}
  c^{\rm ss}_{\ell}\big(\r; S(t)\big)
  = \sum_{\text{cells }\alpha}c^{(1)}_{\ell}(\r-\r^\alpha,\z^\alpha) 
\end{equation}
where $c^{(1)}_{\ell}(\r,\z)$ is the concentration profile produced by
a single isolated cell at position $\r=\bm{0}$, given by the
solution to
\begin{equation} 0 =
    D_{\ell}\nabla_{\r}^2c^{(1)}_{\ell}(\r, \z) -
    d_{\ell}c^{(1)}_{\ell}(\r,\z) + \lambda_{\ell}(\r,\z)
    \label{eq:c1}
\end{equation}
with appropriate boundary conditions.

\subsection{Well-mixed population}
\label{sec:mixed}
In a further approximation, that may apply to \textit{e.g.},
well-mixed populations of stem cells that differentiate in
microenvironments, spatial distributions of cells and signaling
molecules can both be treated as spatially averaged quantities.
Hence, Eq.~\ref{movement} can be ignored and $c_{\ell}$ is independent
of $\r$. If the dynamics of $c_{\ell}$ are given by
Eq.~\ref{lambda_dynamics} with reflecting boundary conditions, then
Eq.~\ref{lambda_dynamics} can be replaced by an ODE for the mean
concentration $\bar{c}(t) = (1/V)\int c_{\ell}(\r)\dd \r$ over the
closed volume $V$
\begin{equation}\label{mixed_lambda}
  \frac{\dd \bar{c}_{\ell}}{\dd t}
  = \sum_{\text{cells }\alpha}\bar{\lambda}_{\ell}(\z^\alpha) - d_{\ell}\bar{c}_{\ell}
\end{equation}
where $\bar{\lambda}_{\ell}(\z)=(1/V)\int \lambda_{\ell}(\r,\z)\dd \r$
is the spatially averaged excretion rate of each of the uniformly
distributed cells.  If signaling molecules reach steady state rapidly,
then the signaling molecule concentrations are

\begin{equation}\label{lambda_mixed_fast}
  c^{\rm ss}_{\ell} = \frac{1}{d_{\ell}}\sum_{\text{cells }\alpha}
  \bar{\lambda}_{\ell}(\z^\alpha).
\end{equation}

\subsection{Fast $\z$ dynamics and discrete cell types}
\label{discrete_types}
For fast cell state dynamics, the state $\z^\alpha$ of cell $\alpha$
rapidly reaches an attractor of $\bm{F}(\z^\alpha;\c^\alpha)$ but can
make stochastic transitions to different attractors. Hence, we can
approximate the continuous state space by the discrete set of
attractors, labeled $q=1,...,Q$. Such a discrete approximation of an
otherwise continuous or highly granular set of cell states has been
shown to be useful in many developmental contexts such as embryonic
stem cell differentiation \cite{DISCRETE,XING_ELIFE,XING_PRXL}.  The
Waddington vector field $\bm{F}(\z^\alpha; \c^\alpha)$ is then
replaced by a matrix of jump rates, $F_{q,q'}(\c^\alpha)$, specifying
the jump rate from $q$ to $q'$. These jump rates can, in principle, be
estimated numerically via simulations of
Eq.~\ref{x_dynamics}. Alternatively, if $\bm{F}(\z^\alpha; \c^\alpha)$
is the gradient of a Waddington landscape, then Kramers' rate formula
can be used to approximate the jump rates \cite{hanggi1990reaction}.

The division rate at which cell $\alpha$ in attractor $q$ divides into
daughter cells at attractor states $q',q''$ with positional
displacements $\Delta \r',\Delta\r''$ is denoted
$\tilde{\beta}_{q,q',q''}(\c; \Delta \r',\Delta\r'')$. Just as for
Eq.~\ref{birth}, $\tilde{\beta}_{q,q',q''}(\c; \Delta \r',\Delta\r'')$
is invariant under exchange of $q'$ and $q''$, as well as under
exchange of $\Delta \r'$ and $\Delta \r''$. The diagonal elements
$\tilde\beta_{q,q,q}$ correspond to symmetric division in which both
daughters have the same state as the mother, whereas elements with
$q'\neq q''$ correspond to asymmetric division. Letting $\z_q$ denote
the attractor state $q$, these division rates are given by
$\tilde{\beta}_{q,q',q''}(\c; \Delta
\r',\Delta\r'')=\tilde{\beta}(\z_q;\c;\z_{q'},\z_{q''},\Delta
\r',\Delta \r'')$. The death rate in attractor $q$ is
$\mu_q(c)=\mu(\z_q; \c)$.
  
\subsection{Unifying the assumptions}
\label{all}

If all three of the above simplifying approximations (fast
equilibration of signaling molecules, well-mixed population, and
discrete cell types) hold, our model simplifies significantly. Let the
vector $\n\equiv(n_1,...,n_Q)$ specify the cell populations in the
tissue, where $n_q$ is the number of cells of type $q$. Let
$\beta_{q,q',q''}(\c)\equiv\int \tilde\beta_{q,q',q''}(\c; \Delta
\r',\Delta\r'')\dd\Delta\r' \dd\Delta\r''$ be the division rate of a
mother of type $q$ into daughters of type $q', q''$. Signaling
concentrations $\c(\n)$ are functions of only $\n$, and are determined
by Eq.~\ref{lambda_mixed_fast}.

The probability $P(\n,t)$ of having a tissue state $\n$ at time $t$
evolves accord to the following master equation:
\begin{widetext}
\begin{equation}
  \begin{aligned}
  \label{reduced_master}
  \partial_t P(\n,t) = & 
  \sum_{q,q',q''} \Big[ (n_q+1)\beta_{q,q',q''}\big(\c(\n_{q+,q'-,q''-})\big)
  P(\n_{q+,q'-,q''-})
  - n_q\beta_{q,q',q''}\big(\c(\n)\big) P(\n) \Big] \\
  \: &  + \sum_{q,q'} \Big[ (n_q+1)F_{q,q'}\big(\c(\n_{q+,q'-})\big)P(\n_{q+,q'-}) 
  - n_q F_{q,q'}\big(\c(\n)\big) P(\n) \Big] \\
  & + \sum_q \Big[ (n_q+1)\mu_q\big(\c(\n_{q+})\big)P(\n_{q+})
  - n_q\mu_q\big(\c(\n)\big)P(\n) \Big]
  \end{aligned}
\end{equation}
\end{widetext}
In analogy with the bookkeeping notation used in
\cite{Xia_JPA,Xia_PRE}, $\n_{q+,q'-,q''-}$ denotes a system state
with, compared to system state $\n$, one more type $q$ cell and one
fewer cell of type $q'$ and one fewer cell of type $q''$. Hence,
$\n_{q+,q'-,q''-}$ is the system state that converts into state $\n$
when a mother cell of type $q$ divides into daughter cells of type
$q'$ and $q''$. Similarly, $\n_{q+,q'-}$ denotes a system state with
one more type $q$ cell and one fewer type $q'$ cell relative to system
state $\n$. $\n_{q+}$ denotes a system state with one more type $q$
cell than in system state $\n$.
%
%
%
Equation~\ref{reduced_master} represents a master equation for the
probability density for populations of cells of multiple types and is
a simpler form of master equations studied in
\cite{BDI,Xia_JPA,Xia_PRE}.
%
%
For small populations and few accessible cell types, solving this
linear master equation or simulating the process is computationally
feasible.

\subsection{Developmental Fitness landscape}
Proliferation of certain cell types over others due to differing
division and death rates is a crucial aspect of
development. Differential proliferation is also key to evolutionary
dynamics, where genotypes that induce high net growth rates eventually
comprise larger portions of the overall population. In evolutionary
dynamics, preferential growth is described by a fitness landscape,
typically defined as a function which specifies the net growth
(division minus death) rate $\phi_g$ of a cell or organism given its
genotype $g$. Here, we adapt this concept to the context of cell
development to define an epigenetic fitness landscape which
specifies which cell types proliferate in the developing tissue.

A key difference between evolutionary dynamics (or speciation) and
development is that, in a developing tissue, all cells have the same
genotype (aside from rare mutations) and differences in cells'
division rates are due to differences in their cell type, determined
by the internal state $\z$. We define the developmental fitness of
cell type $q$, denoted $\phi_q(\n)$, as the net proliferation rate of
type $q$ cells, absent production of type $q$ cells through division
of other cell types or stochastic jumps from other cell
types. $\phi_q(\n)$ depends on the tissue population state $\n$ due to
cell-cell interactions via signaling molecules. In defining
$\phi_q(\n)$, we adopt the assumptions from the previous sections,
namely, fast equilibration of signaling molecules, well-mixed
population, and discrete cell types. Within our framework, we can
define fitness mathematically as
\begin{equation}
  \begin{aligned}
    \phi_q(\n) = & \beta_{q,q,q}\big(\c(\n)\big)
    - \beta_{q,\rm lost}\big(\c(\n)\big)\\
    \: & \qquad\quad -  F_{q,\rm lost}\big(\c(\n)\big) - \mu_q\big(\c(\n)\big)
\end{aligned}
  \label{fitness}
\end{equation}
where $\beta_{q,q,q}(\c)$ is the rate of symmetric division in which a
mother cell of type $q$ divides into two daughter cells of type $q$.
$\beta_{q,\text{lost}}(\c)\equiv\sum_{q'\neq q,q''\neq
  q}\beta_{q,q',q''}(\c)$ is the rate at which a mother cell of type
$q$ divides into daughter cells, both of whose cell types differ from
that of the mother cell. Note that asymmetric division, where a cell
of type $q$ divides into daughters of type $q$ and $q'\neq q$, leaves
the population of type $q$ cells unchanged, and therefore does not
factor into the developmental fitness landscape. $F_{q,\text{lost}}(\c)
= \sum_{q'\neq q} F_{q,q'}(\c)$ is the jump rate of type $q$ cells to a
different type.

The following subsection shows how, in the limit of large tissue size,
the developmental fitness landscape contributes to tissue dynamics
through a simple equation.

\subsection{Deterministic limit}
\label{V_to_inf}

In the limit of large tissue size, Eq.~\ref{reduced_master} predicts
that cell densities in the tissue follow deterministic dynamics. This
limit is analogous to how deterministic mass action equations (ODEs)
emerge from a more microscopic chemical master equation description in
the limit of large system size. Recall that Eq.~\ref{reduced_master}
describes the population within a region of fixed volume, and denote
this volume by $V_0$. Then, $\boldsymbol{\rho}\equiv\n/V_0$ are the
cell type densities in the tissue described by
Eq.~\ref{reduced_master}. Now, consider a system with volume $V>V_0$
and take the limit that $V\to\infty$ while $\boldsymbol{\rho}$ remains
fixed. The mass-action dynamics in this limit are described by
\begin{equation}
  \label{deterministic}
  \frac{\dd \boldsymbol{\rho}}{\dd t}
  = \text{diag}(\bm\phi(\bm\rho V_0))\bm\rho + \bm m(\bm\rho V_0) \bm\rho
\end{equation}
where $\boldsymbol\phi(\n)=(\phi_1(\n),...,\phi_Q(\n))$ is the
developmental fitness landscape defined in the previous section, and
$\text{diag}(\bm\phi(\bm\rho V_0))$ denotes the $Q\times Q$ matrix
with the elements of $\boldsymbol\phi(\boldsymbol\rho V_0)$ on the
diagonal. $\boldsymbol m(\boldsymbol\rho V_0)$ is a zero-diagonal
matrix which contains the jump rates and asymmetric division rates
through which cells of one type produce cells of another type. The
matrix elements are explicitly

\begin{equation}
  \begin{aligned}
    m_{q,q'}(\n) = & F_{q',q}\big(\c(\n)\big)
    + 2\beta_{q',q,q}\big(\c(\n)\big) \\
    \: & \qquad\qquad  + 2\!\sum_{q''\neq q}\beta_{q',q,q''}\big(\c(\n)\big).
    \end{aligned}
\end{equation}
The factor of 2 in the third term arises from the symmetry of
$\beta_{q',q,''}$ with respect to exchange of the final two indices.

Deterministic dynamical models of interacting population dynamics, as
could be described by Eq.~\ref{deterministic}, have been used to model
aspects of development \cite{jain2024cell}. Note that the stochastic
fluctuations in the dynamics of $\bm\rho$ are proportional to
$(V/V_0)^{-1/2}$, which vanishes as $V\to\infty$. For large but finite
$V$, the stochastic component of the dynamics can be obtained with the
van Kampen system size expansion \cite{gardiner1985handbook}.

\section{Waddington's Landscape vs Waddington's vector field}\label{landscapes}

From among the various interpretations of Waddington's landscape that
have been studied. The most common interpretation that has emerged is
that the vector field $\bm F(\z;\c)$ in Eq.~\ref{x_dynamics} is a
negative gradient of a function $U(\z;\c)$,

\begin{equation}\label{grad_dynamics}
  \bm{F}(\z;\c) = -\nabla_{\z} U(\z;\c),
\end{equation}
where $U(\z;\c)$ is called a Waddington landscape
\cite{schiebinger2021reconstructing,shi2019quantifying,weinreb2018fundamental,coomer2022noise}. We
call a stochastic dynamical system of this form, where the
deterministic part is the negative gradient of a potential landscape,
a \textit{gradient system}. If the noise in Eq.~\ref{x_dynamics} is
additive (meaning $\sigma$ is independent of $\z$), then $U(\z;\c)$ can
be reconstructed from experimental measurement of the steady state
distribution of cell states \textit{at fixed signaling molecule
  concentrations}, $p_{\rm ss}(\z;\c)$, via
\cite{schiebinger2021reconstructing}
\begin{equation}
  \label{reconstructed}
  U(\z;\c) = -\log p_\text{ss}(\z;\c).
\end{equation}

The validity and generality of this Waddington landscape framework has
been a topic of considerable discussion and disagreement. On one
hand, it has been pointed out that realistic dynamics of protein and
mRNA concentrations are not gradient systems, but rather have
\textit{curl} (or the higher-dimensional generalization of curl)
\cite{wang2011quantifying, qiu2022mapping,shi2022energy}. On the other
hand, it has been noted that for nearly all vector fields $\bm{F}$
without limit cycles (specifically, for Morse-Smale systems), the
vector field $\bm F$ can be converted to the negative gradient of a
potential landscape with a coordinate transformation
\cite{rand2021geometry}. Hence, a very broad class of models,
including most models of protein and mRNA dynamics, can be expressed
as gradient systems under a suitable choice of coordinates. This
suggests that the Waddington landscape picture is highly general.

However, noise has a crucial effect which significantly hinders the
generality of the landscape picture. Eq.~\ref{reconstructed} is valid
only for additive noise, yet the noise in the dynamics of protein and
mRNA concentrations is typically non-additive
\cite{coomer2022noise}. It was recently demonstrated how the
mis-application of Eq.~\ref{reconstructed} to systems with
non-additive noise can `add' or `delete' attractors, meaning that the
reconstructed $U(\z;\c)$ from Eq.~\ref{reconstructed} may have
attractors not present in the true dynamics, or may not have
attractors that are present in the true dynamics
\cite{coomer2022noise}. We demonstrate a similar effect for systems
with curl: if Eq.~\ref{reconstructed} is mis-applied to a system with
curl and with additive noise, the reconstruction of $U(\z;\c)$ can also
`add' or `delete' attractors (see appendix A). In fact, these two
mis-applications of Eq.~\ref{reconstructed}, to systems with
non-additive noise and to systems with curl, are two sides of the same
coin: if a non-gradient system with additive noise is converted via
coordinate transformation to a gradient system, the noise will no
longer be additive, so Eq.~\ref{reconstructed} will be
invalid. Conversely, if a gradient system with non-additive noise is
converted to a system with additive noise via coordinate
transformation, the dynamics will no longer be gradient, so
Eq.~\ref{grad_dynamics} will be invalid. In short, one can change
coordinates to remove curl or non-additive noise, but not both. Hence,
coordinate transformations cannot allow a system to simultaneously
satisfy both Eqs.~\ref{grad_dynamics} and \ref{reconstructed}, so the
Waddington landscape is not a general framework for studying gene
regulatory dynamics.

We advocate for working directly with the vector field $\bm F(\z;\c)$
that describes gene regulatory dynamics, which we call
\textit{Waddington's vector field}, regardless of whether it is the
negative gradient of a landscape. Importantly, the key conceptual
clarity offered by Waddington's landscape--that cell states are drawn
into attractors governed by gene regulatory networks--remains valid in
a `Waddington vector field' picture. While some authors propose
decomposing the vector field into gradient and curl parts
\cite{wang2008potential,wang2011quantifying,shi2022energy,lei2023mathematical},
this decomposition can be misleading because basins of attraction of
the gradient part may not contain any attractor of $\bm F(\z;\c)$, and
we see no advantage to this decomposition. The Waddington vector field
picture is, we argue, a clearer and mathematically precise framework
to describe the gene regulatory dynamics in cell development.

\section{Example 1: Stem cell differentiation in a
  well-mixed microenvironment}\label{ex1}

Here, we present a model of tissue development which is derived by
augmenting a prototypical model of Waddington's landscape studied in
\cite{huang2007bifurcation,wang2011quantifying,qiu2022mapping}. In our
model, $\bm{F}(\z;\c)$ describes a hypothetical gene regulatory
circuit involving two proteins, ${\rm A}$ and ${\rm B}$. $\z=(z_{\rm
  A}, z_{\rm B})$ specifies the concentrations of these two
proteins. $\bm{F}(\z;\c)$ has three attractors corresponding to three
cell types, which, for reasons that will become clear, we call
\textit{stem cells} $\mathcal{S}$, \textit{triggered stem cells}
$\mathcal{S}^{*}$, and \textit{differentiated cells}
$\mathcal{D}$. Cells interact via two signaling molecules with
concentrations $\c=(c_1, c_2)$. These interactions induce stem cells to
differentiate in a coordinated way, so that a stable population
(\textit{i.e.} a tissue) of differentiated cells is produced and
maintained. Hence, while the previous models in
\cite{huang2007bifurcation,wang2011quantifying,qiu2022mapping}
describe only single-cell differentiation, our model describes the
coordinated development of tissue. After analyzing the dynamics of the
full model, we show how the model can be simplified into a simple
system of ODEs using the approximations developed in
section~\ref{assumptions}.

\begin{figure*}
	\centering
	\includegraphics[width=0.94\linewidth]{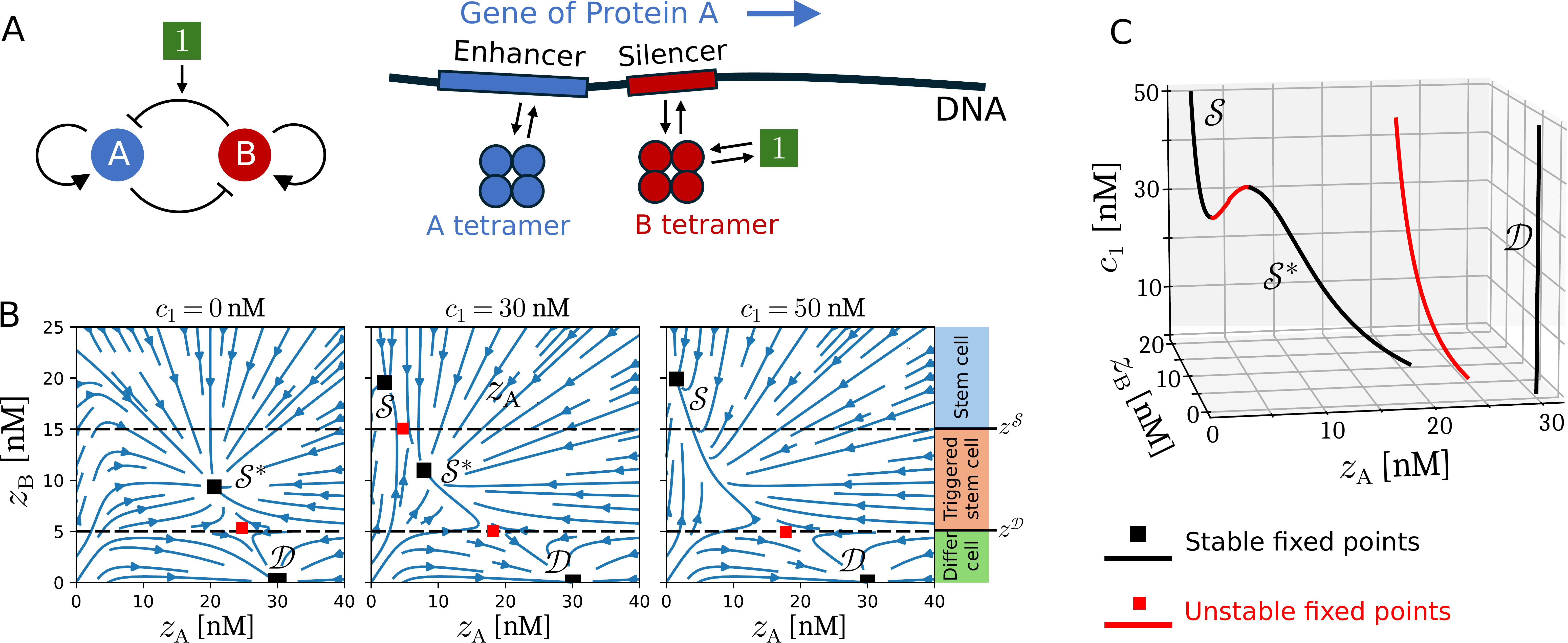}
	\caption{Example 1: Model of regulated stem cell
          differentiation. (A) Gene regulatory circuit. Left:
          Schematic of circuit showing upregulation (arrows), down
          regulation (flat arrows), and the effect of signaling
          molecule 1 concentration $c_1$ on amplification of the
          downregulation of ${\rm A}$ by ${\rm B}$. Right: Diagram of
          molecular mechanisms. Black arrows indicate molecular
          binding and unbinding of the A and B tetramers to the
          enhancer and silencer, and of signaling molecule 1 (green
          square) to B tetramer. Signaling molecule 2 controls cell
          proliferation and is not featured in this picture of
          individual-cell gene regulation dynamics. (B) Vector field
          $\bm F(\z;\c)$ at three values of $c_1$. Attractors (black
          squares) are labelled by cell type: stem cells
          $\mathcal{S}$, triggered stem cells $\mathcal{S}^*$, and
          differentiated cells $\mathcal{D}$. Unstable fixed points
          indicated by red squares. Black dashed lines at
          $z_\text{B}=z^\mathcal{S}$ and $z^\mathcal{D}$ denote the
          boundary between stem cells and triggered stem, and between
          triggered stem cells and differentiated cells,
          respectively. (C) Fixed points of $\bm{F}(\z;\c)$ as a
          function of $c_1$. Two saddle-node bifurcations result in
          the gain/loss of attractors $\mathcal{S}$ and
          $\mathcal{S}^*$.}
	\label{fig:F}
\end{figure*}

The gene regulatory circuit that governs the dynamics of $\z$ is
illustrated in Fig.~\ref{fig:F}A. Each gene (\textit{A} and
\textit{B}) upregulates itself and downregulate the other. We make the
common quasi-steady state assumption that mRNA transcription and
degradation are fast relative to protein translation and degradation,
so the state and dynamics are described in terms of only the protein
concentrations $\z=(z_{\rm A}, z_{\rm B})$. Signaling molecule ``1''
influences expression of gene \textit{A}, but not gene
\textit{B}. Specifically, B tetramer (which downregulates gene
\textit{A}) can form a complex with signaling molecule 1, and this
complex prevents any expression of A when bound to the silencer of
A. This binding scheme gives rise to a Waddington vector field, shown
in Fig.~\ref{fig:F}B, whose components are
  
\begin{equation}
    \begin{aligned}
    F_m(\z;\c) = & r_{0,m}\big(1-E_m(z_m)\big)\big(1-S_m(z_{m'})\big)\\
  \: & \,\,+ r_{E,m}E_m(z_m)\big(1-S_m(z_{m'})\big)\\ 
  \: & \,\,+ r_{S,m}(c_1)\big(1-E_m(z_m)\big)S_m(z_{m'})\\
  \: & \,\,+ r_{ES,m}(c_1)E_m(z_m)S_m(z_{m'}) - k_{m}z_{m}\\
\end{aligned}
    \label{ex1_F}
  \end{equation}
where $m,m'=A,B$ or $B,A$. $E_m(z_m)$ and $S_m(z_{m'})$ are,
respectively, the probabilities that the enhancer and silencer of gene
$m$ have a bound tetramer, and are modeled by Hill functions:

\begin{equation}\label{Enhancer_silencer}
  E_m(z_m) = \frac{z_m^4}{K_{E,m}^4 + z_m^4},\,\,\,
	S_m(z_{m'}) = \frac{z_{m'}^4}{K_{S,m}^4 + z_{m'}^4}.
\end{equation}
The rates $r_{0,m}$, $r_{E,m}$, $r_{S,m}(c_1)$, and $r_{ES,m}(c_1)$
are the mean expression rates of gene $m$ with, respectively, neither
enhancer nor silencer occupied, only enhancer occupied, only silencer occupied,
and both enhancer and silencer occupied. The rates $r_{S,A}(c_1)$ and
$r_{ES,A}(c_1)$ have a Michaelis-Menten type dependence on $c_1$:

\begin{equation}\label{B1_binding}
  r_{S,A}(c_1) = \frac{r_{S,A}(0) \gamma}{\gamma + c_1},\quad 
  r_{ES,A}(c_1) = \frac{r_{ES,A}(0) \gamma}{\gamma + c_1}.
\end{equation}
The factor $\gamma / (\gamma + c_1)$ is the probability that a given B
tetramer is \textit{not} bound to a signaling molecule. All other rate
parameters, $r_{0,m}$, $r_{E,m}$, $r_{S,B}$, $r_{ES,B}$, $r_{S,A}(0)$,
and $r_{ES,A}(0)$ are constants (listed in Table 1, along with all
other model parameters). Fig.~ \ref{fig:F}B shows the Waddington
vector field for three values of $c_1$: $c_1=0$ (left), $c_1=30$nM
(middle), and $c_1=50$nM (right). For a particular choice of
parameters, $\bm{F}(\z;\c)$ reduces to the model in
\cite{huang2007bifurcation,wang2011quantifying}.

The key role of signaling molecule $1$ is to regulate how stem cells
are triggered for differentiation. Fig.~\ref{fig:F}B shows the
attractors (black squares) for the three cell types ($\mathcal{S}$,
$\mathcal{S}^{*}$, and $\mathcal{D}$). At intermediate $c_1$, stable
attractors exist for both $\mathcal{S}$ and $\mathcal{S}^*$, but when
$c_1$ drops sufficiently low, the $\mathcal{S}$ attractor disappears
and stem cells move toward the triggered state
$\mathcal{S}^{*}$. Hence, low $c_1$ triggers stem cells for
differentiation, and as we will see below, these triggered stem cells
undergo asymmetric division to produce differentiated cells. On the
other hand, very high $c_1$ reverses the triggering (the
$\mathcal{S}^*$ attractor disappears, so triggered stem cells revert
to untriggered stem cells). Fig.~\ref{fig:F}C shows how the locations
of the attractors (black) and unstable fixed points that separate the
attractors (red) vary continuously with $c_1$. The unstable fixed
point that separates the $\mathcal{S}$ and $\mathcal{S}^{*}$
attractors merges with $\mathcal{S}$ at low $c_1$, but merges with
$\mathcal{S}^*$ at high $c_1$. These mergers of fixed points are
saddle-node bifurcations.

Cell division rates depend on both the expression level $z_{\rm B}$ of protein B 
and the concentration $c_2$ of signaling molecule 2. The dependence on
$z_{\rm B}$ gives rise to the defining behavior of the three cell
types: stem cells divide symmetrically (both daughters are type
$\mathcal{S}$), triggered stem cells divide asymmetrically (one daughter
is $\mathcal{S}$ and the other $\mathcal{D}$), and differentiated
cells do not divide. Mathematically, we model the birth rate by
\begin{equation}\label{betas}
\tilde\beta(\z;\z',\z'')= \hspace{-1mm}\begin{cases}
	0 & z_{\rm B} \leq z^{\mathcal{D}} \\[5pt]
        {\small \beta_{\mathcal{S}^{*}}\delta (\z - \z'-\bm{w}(\z))} \\
          \: \quad {\small \times 
  \delta(\z - \z''+\bm{w}(\z))} & z^{\mathcal{S}}\!\geq z_{\rm B}\! > z^{\mathcal{D}} \\[6pt]
  \displaystyle \frac{a_1\delta (\z - \z')\delta(\z - \z'')}{1+a_2\exp(a_3 c_2)} & z_{\rm B}>z^{\mathcal{S}}.
	\end{cases}
\end{equation}
where $z^{\mathcal{D}}< z^{\mathcal{S}}$ are threshold values for
$z_{\rm B}$. Note that we do not include the $\Delta\r',\Delta\r''$
dependencies because this model assumes a well-mixed (spatially
uniform) system. The threshold concentrations $z^{\mathcal{S}}$ and
$z^{\mathcal{D}}$ are indicated by dashed lines in Fig.~\ref{fig:F}B,
and are determined by the (approximate) $z_{\rm B}$ values of the
unstable fixed points of $\bm{F}$.

The vector $\bm{w}(\z)$ induces asymmetric division: when a triggered
stem cell divides, one daughter has a state shifted by $\bm{w}(\z)$
relative to the mother, and the other daughter has state shifted by
$-\bm{w}(\z)$. As a result, one daughter returns to the $\mathcal{S}$
attractor, while the other moves to the $\mathcal{D}$ attractor.  Note
that for $\z$ defined by concentrations, the concentrations $\z'$ and
$\z''$ in the daughter cells are related to that of the mother through
$(\z'+\z'')/2 = \z$.  The death rate $\mu$ for all cells is assumed
constant.
  
In this model, signaling molecule 1 is excreted by differentiated
cells and signaling molecule 2 is excreted by stem cells. We use
Eq.~\ref{lambda_mixed_fast} to model signaling molecule
concentrations, which assumes a well-mixed population and that
concentrations rapidly reach steady state. The excretion rates of
$c_{1}$ and $c_{2}$ are, respectively,
\begin{equation}\label{phi1}
	\bar{\lambda}_1(\z) = \begin{cases}
	0 &  z_{\rm B} > z^{\mathcal{D}}\\
	L_1 &  z_{\rm B} \leq z^{\mathcal{D}}
	\end{cases}
\end{equation}
and
\begin{equation}\label{phi2}
	\bar\lambda_2(\z) = \begin{cases}
	L_2 &  z_{\rm B} > z^{\mathcal{S}}\\
	0 &  z_{\rm B} \leq z^{\mathcal{S}}
	\end{cases}
\end{equation}
Thus, according to Eq.~\ref{lambda_mixed_fast}, in a tissue with
$N_\mathcal{S}$ stem cells and $N_\mathcal{D}$ differentiated cells,
the concentrations are simply proportional to the cell numbers
$n_\mathcal{S}$ and $n_\mathcal{D}$,
\begin{equation}\label{ex1_c}
  (c^{\rm ss}_1, c^{\rm ss}_2) = \Big(n_\mathcal{D}\mfrac{L_1}{d_1},
  n_\mathcal{S}\mfrac{L_2}{d_2}\Big).
\end{equation}

\begin{figure*}
	\centering
	\includegraphics[width=0.92\linewidth]{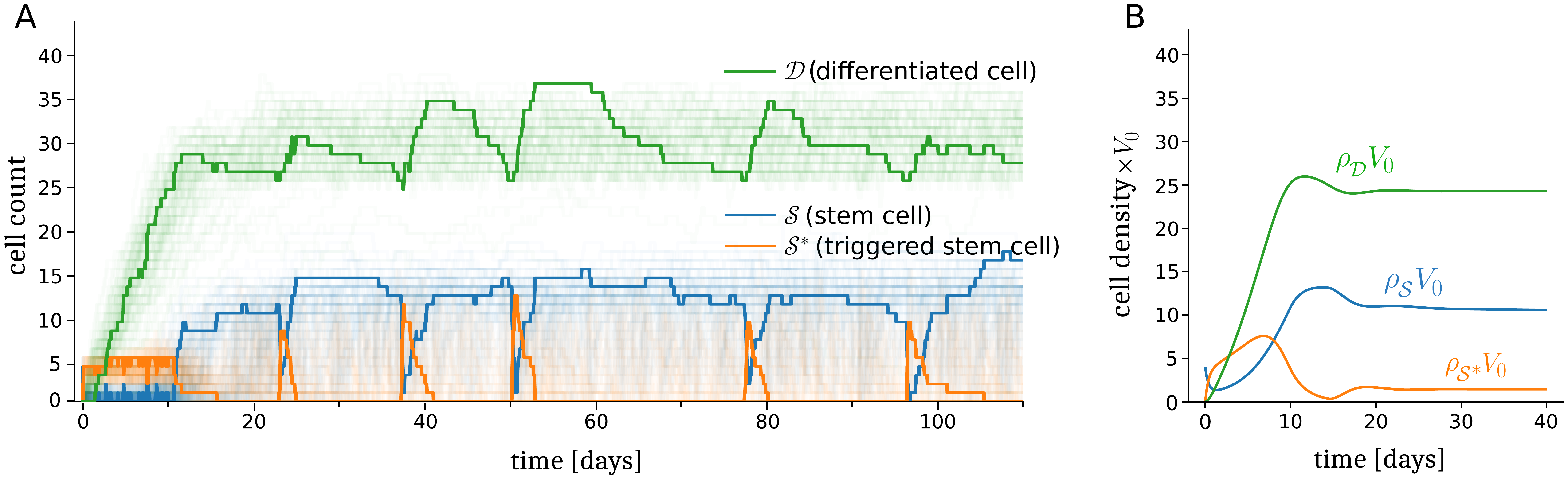}
	\caption{Example 1: Dynamics of cell populations. (A)
          Dynamical simulation starting from an initial condition with
          $N_{\cal S}=4$ and $N_{{\cal S}^*}=N_D=0$. The population
          dynamics exhibit thermostat-like behavior: when $N_{\cal D}$
          falls below 26, stem cells are triggered for asymmetric
          division, which results in $N_{\cal D}$ being
          replenished. One-hundred trajectories were simulated by
          evaluating Eq.~\ref{x_dynamics}, coupled with a discrete
          cell birth-death process, using the standard Euler-Maruyama
          method. One representative trajectory is highlighted, while
          the other ninety-nine are lightly shaded. (B) Evolution of
          cell populations in the deterministic limit. The densities
          are multiplied by the normalizing volume $V_{0}$ to express
          densities on the same scale as particle counts. Note that
          the deterministic trajectories qualitatively reflect the
          overall dynamics shown in (A).}
	\label{fig:F2}
\end{figure*}

The signaling molecules' influence on gene dynamics and birth rates
gives rise to robust tissue development: from a wide range of initial
conditions, the system dynamically approaches a stable tissue state
with regulated relative cell type populations, as illustrated in
Fig.~\ref{fig:F2}A. The two signaling molecules govern the two key
regulatory mechanisms: \textit{(i)} the stem cell population,
$n_\mathcal{S}$, is self-regulated due to the dependence of stem cell
division on $c_2$. When $n_\mathcal{S}$ is low (high), $c_2$ is low
(high), so stem cell division rate is high (low), replenishing
(diminishing) $n_\mathcal{S}$ towards the stable, developed state;
\textit{(ii)} the dependence of $\bm{F}(\z;\c)$ on $c_1$ leads to a
thermostat-like behavior of stem cell triggering. When $n_\mathcal{D}$
is low (high), $c_1$ is low (high), so more (less) stem cells are
triggered for asymmetric division. This thermostat-like behavior is
apparent from Fig.~\ref{fig:F2}A; whenever $n_\mathcal{D}$ drops below
a threshold, stem cells are triggered, rapidly replenishing
$n_\mathcal{D}$. In Fig.~\ref{fig:F2}A specifically, an initial
population of only four stem cells develops into the stable tissue
state.

\subsection{Deterministic limit}

The deterministic limit described in section~\ref{V_to_inf} can be
applied to obtain a highly simplified form of the model. Denote the
cell densities for the three cell types as $\bm\rho =
(\rho_\mathcal{S},\rho_{\mathcal{S}^*},\rho_\mathcal{D})$. The
developmental fitness landscape is

\begin{equation}
  \bm\phi(\n) = \big(\beta_\mathcal{S}(n_\mathcal{S})-
  F_{\mathcal{D},\mathcal{S}}(n_\mathcal{D}){-}\mu,\,\,
  -\beta_{\mathcal{S}^*}{-}\mu, \,\, -\mu \big)
\end{equation}
where $\beta_\mathcal{S}(n_\mathcal{S})\equiv a_1/(1+a_2\exp{(a_3
  n_\mathcal{S}L_2/d_2)})$, as determined by Eqs.~\ref{betas} and
\ref{ex1_c}. $F_{\mathcal{D},\mathcal{S}}(n_\mathcal{D})$ is the rate
at which stem cells are
triggered. $F_{\mathcal{D},\mathcal{S}}(n_\mathcal{D})$ can be
estimated numerically by simulating the gene dynamics, and it is well
approximated by a step function, equal to 0 for $n_\mathcal{D}>25$ and
greater than 0 for $n_\mathcal{D}\leq 25$. The deterministic equations
given by Eq.~\ref{deterministic} written out explicitly are

\begin{equation}
\begin{aligned}
  \frac{\dd\rho_\mathcal{S}}{\dd t} & = \big(\beta_\mathcal{S}(\rho_\mathcal{S}V_0)-
  F_{\mathcal{D},\mathcal{S}}(\rho_\mathcal{D}V_0)-\mu \big) \rho_\mathcal{S}
  + \beta_{\mathcal{S}^*}\rho_{\mathcal{S}^*} \\
  \frac{\dd\rho_{\mathcal{S}^*}}{\dd t} & =
  F_{\mathcal{D},\mathcal{S}}(\rho_\mathcal{D}V_0)\rho_\mathcal{S}
  -\big(\beta_{\mathcal{S}^*}+\mu\big)\rho_{\mathcal{S}^*}\\
  \frac{\dd\rho_\mathcal{D}}{\dd t} & = \beta_{\mathcal{S}^*}
  \rho_{\mathcal{S}^*} - \mu \rho_\mathcal{D}.
\end{aligned}
\label{Wad_pop}
\end{equation}
Fig.~\ref{fig:F2}B plots the solutions of these equations for the same
initial conditions as in Fig.~\ref{fig:F}D. The transient
thermostat-like behavior is lost, but the longer timescale results are
qualitatively similar.

\section{Example 2: Spatial model inspired by planaria}\label{ex2}

Here, we present a model of a spatially-structured multicellular
organism which develops from a single cell. Additionally, if the
developed organism is cut into pieces, each piece will regenerate into
a new fully-developed organism (a phenomenon called fissiparous
reproduction). This model is inspired by planaria worms, which exhibit
a remarkable capacity for fissiparous reproduction: a fragment of a
planarian as small as 1/279th of the full worm can regenerate into a
new worm \cite{morgan1898experimental,lobo2012modeling}.

In this model, cell state is described by the expression of a single
gene, denoted gene A, whose expression is governed by a bistable
regulatory circuit. The state $z$ is the concentration of protein A
(we assume fast mRNA dynamics, as in example 1). Two stable states
corresponding to two values of $z$ define two cell types: exterior
cells (which have low $z$) and interior cells (which have high
$z$). Cells excrete two signaling molecules, one which governs the $z$
dynamics and another which governs cell division rate. The cells live
and migrate on a 2D substrate ($\r\in\mathbb{R}^2$), and they exert
forces on one another which govern their movement.

\begin{figure*}
	\centering
	\includegraphics[width=0.86\linewidth]{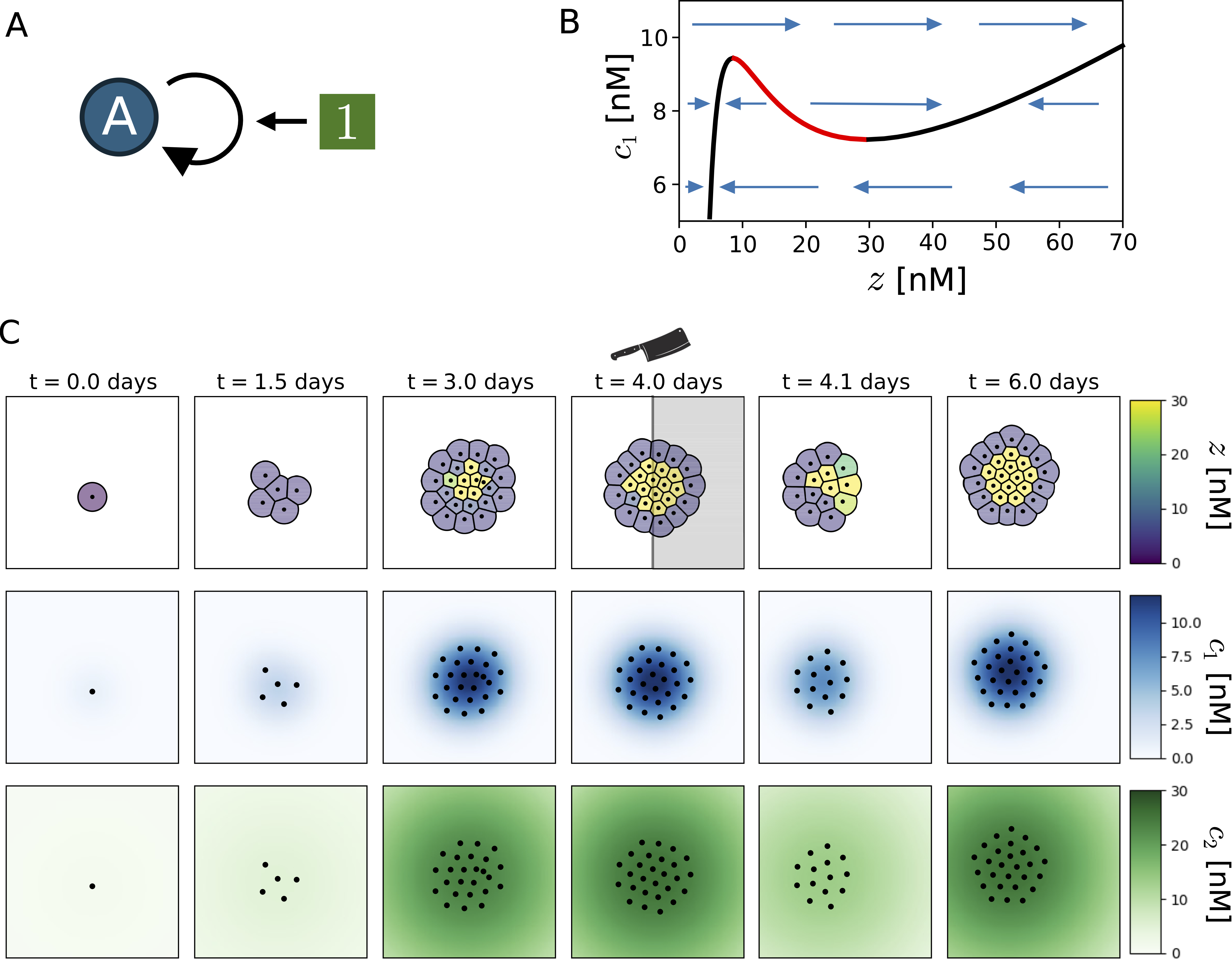}
	\caption{Example 2: Model of a spatially structured organism
          with fissiparous reproduction. (A) Gene regulatory circuit,
          where gene A upregulates itself in the presence of signaling
          molecule 1. (B) Waddington vector field (blue arrows),
          attractors (black curves) and unstable fixed points (red
          curve) as a function of $c_1$. The lefthand attractor
          represents exterior cells and the righthand attractor
          represents interior cells. (C) Simulation of the dynamics,
          showing a single cell developing into a stable developed
          organism (day 0 through 4). At time $t=4.0$ days, the
          organism is bisected and the left half is removed. The
          organism then regrows (4.1 to 6.0 days). Top row: cell
          position (black dots) and the state level $z$ given by the
          color shading. Middle and bottom rows depict the
          corresponding concentrations $c_1$ and $c_2$, respectively.
         \label{fig:ex2}}
\end{figure*}

The gene regulatory circuit is illustrated in
Fig.~\ref{fig:ex2}A. Gene A upregulates itself when signaling molecule
1 is present. Specifically, protein A can bind to signaling molecule
1, and this heterodimer binds to the enhancer of gene A, increasing
the expression of protein A. The Waddington vector field thus takes
the form
\begin{equation}\label{ex2_F}
	F(z;c) = r_0 + \frac{r_1 c_1 z^2}{K^2 + z^2} - b z.
\end{equation}
The values of the parameters $r_0$, $r_1$, $K$, and $b$, as well of
the parameters introduced in the following equations, are given in
Table 1.

Fig.~\ref{fig:ex2}B shows the dynamics (blue arrows) as well as the
locations of the attractors (black) and unstable fixed points (red) of
$F(z; c)$, for values of $c_1$ between 4 and 12 nM. The attractor at
low (high) $z$, which corresponds to exterior (interior) cells, merges
with the unstable fixed point at high (low) $c_1$ via a saddle-node
bifurcation; when $c_1$ reaches this bifurcation, exterior (interior)
cells convert to interior (exterior) cells. In this way, cell type is
regulated by $c_1$.

Cell movement is described by Eq.~\ref{movement} with
\begin{equation}
  \label{ex2_move}
	\bm{g}(\r^\alpha, S) = \sum_{\alpha'\neq \alpha} \bm{v}(\r^\alpha - \r^{\alpha'})
\end{equation}
where $\bm{v}(\r) = v_{0} (\frac{r_\text{min}^2}{r^3}
-\frac{r_\text{min}}{r^2})\bm{\hat{r}}$, where $\bm{\hat{r}}$ is the
unit vector in the $\bm r$ direction and $v_{0}$ is a parameter that
sets the velocity scale (see Table 1).  We assume the motion is
deterministic, so $\eta=0$ in Eq.~\ref{movement}. The velocity
$\bm{v}(\r)$ moves cells separated by a distance greater than
$r_\text{min}$ closer together (\textit{e.g.}, response to an
attractive force), but moves cells separated by less than
$r_\text{min}$ further apart, corresponding to a close-ranged
repulsive force. Hence, the cells that comprise the tissue adhere
together with an average spacing of approximately $r_\text{min}$.

The spatial profile of signaling molecule concentrations is key to the
regulation of cell number and cell type. We assume rapid equilibration
of signaling molecule concentrations (section \ref{fast_c}), with
$c_\ell^\text{ss}$ given by Eq.~\ref{c_ss} with $z$-independent
$c^{(1)}_\ell(\r)= C_\ell\exp\big(-r^2/(2s_\ell^2)\big)$ for
$\ell=1,2$ and $r=|\r|$. While the concentration field
$c^{(1)}_\ell(\r)$ due to a single cell can be computed from
Eq.~\ref{eq:c1}, it can also be well-approximated by a Gaussian spread
function with appropriately chosen variance $s_{\ell}^{2}$ (see Table
1). The concentration $c_1$ informs cells as to whether they are
surrounded by many (high $c_1$) or few (low $c_1$) neighboring
cells. The gene regulatory circuit ensures that cells with many
neighbors become interior cells and cells with fewer neighbors become
exterior cells.

As in Example 1, the concentration of molecule 2, $c_2$, mediates the
total cell population by influencing cell division rates, which we
model by
\begin{equation}\label{ex2_beta}
    \tilde{\beta}(\c;\Delta\r',\Delta\r'') =
    \begin{cases} \beta_0 \delta(\Delta\r'){\cal N}(\Delta\r'') & c_2 < c_2^{*} \\
    0 & c_2 \geq c_2^{*},
    \end{cases}
\end{equation}
where ${\cal N}(\r)\equiv \frac{1}{2\pi\xi^2}e^{-\r^2/2\xi^2}$ is the
2-dimensional Gaussian density function with variance $\xi^2$. Note
that the division rate is independent of $z$.

Fig.~\ref{fig:ex2}C shows a simulation of the model. Starting from a
single cell at $t=0$, the organism grows into a fully-developed
organism in roughly four days. For approximately the first two days,
all cells are of the exterior type (purple). After the organism grows
sufficiently large that $c_1$ concentration reaches a sufficient
level, conversion to interior-type cells (yellow) occurs. In
Fig.~\ref{fig:ex2}C, the top row indicates the positions and
$z$-states of cells, the middle row shows signaling molecule 1
concentrations, and the bottom row shows signaling molecule 2
concentrations.
  
If the fully developed organism is bisected, each half regrows. In the
simulation shown in Fig.~\ref{fig:ex2}C, the organism is bisected at
time $t=4.0$ days along the dotted line shown in the figure, and the
righthand portion is removed. The remaining lefthand portion then
regrows into a new fully-developed organism in approximately 2
days. These results illustrate an extremely robust developmental
process.

\begin{table*}
    \caption{\label{tab:combined_parameters} Model Parameters for Examples 1 and 2.}
    \begin{ruledtabular}
    \begin{tabular}{crll}
        \textbf{Parameter} & \textbf{Value} & \textbf{[units]} & \textbf{Description} \\ \hline
        \multicolumn{3}{l}{\textit{Example 1 Model Parameters}} \\ \hline
        $a_1$ & 1 &[day$^{-1}$] & Max. stem cell division rate (Eq.~\ref{betas}) \\ 
        $a_2$ & 0.15& [unitless] & Parameter in Eq.~\ref{betas} \\ 
        $a_3$ & 0.5& [unitless] & Parameter in Eq.~\ref{betas} \\ 
        $\beta_{\mathcal{S}^*}$ & 0.6& [day$^{-1}$] & Triggered stem cell division rate (Eq.~\ref{betas}) \\ 
        $\bm{w}(\z)$ & $0.95(z_{\rm A}, -z_{\rm B})$ & $[z]$ & Controls asymmetric division (Eq.~\ref{betas}) \\ 
        $\mu$ & 0.02& [day$^{-1}$] & Death rate \\ 
        $L_1$, $L_2$ & 1& [nM] & Mean excretion rates of signaling molecules (Eqs.~\ref{phi1}, \ref{phi2}) \\ 
        $z^\mathcal{S}$ & 15& [nM] & Threshold value of $z_\text{B}$ between stem cells and triggered stem cells (Eqs.~\ref{betas}, \ref{phi2}) \\ 
        $z^\mathcal{D}$ & 5& [nM] & Threshold value of $z_\text{B}$ between triggered stem and differentiated cells (Eqs.~\ref{betas}, \ref{phi1}) \\ 
        $K_{E,m}$ & 5& [nM] & Equilibrium const. for tetramerization and binding of $m$=A, B to Enhancer (Eq.~\ref{Enhancer_silencer}) \\ 
        $K_{S,m}$ & 5& [nM] & Equilibrium const. for tetramerization and binding of $m$=A, B to Silencer (Eq.~\ref{Enhancer_silencer}) \\ 
        $\gamma$ & 20& [nM] & Equilibrium const. for signaling molecule 1 binding to B tetramer (Eq.~\ref{B1_binding}) \\ 
        $r_{0,A}$ & 500& [nM/day] & Expression rate of A when A enhancer and silencer of A are unoccupied (Eq.~\ref{ex1_F}) \\ 
        $r_{0,B}$ & 500& [nM/day] & Expression rate of B when B enhancer and silencer of B are unoccupied (Eq.~\ref{ex1_F}) \\ 
        $r(0)_{S,A}$ & 250& [nM/day] & Expression rate of A when A silencer is occupied and $c_1=0$ (Eq.~\ref{B1_binding}) \\ 
        $r(0)_{ES,A}$ & 1000& [nM/day] & Expression rate of A when A enhancer and silencer are occupied and $c_1=0$ (Eq.~\ref{B1_binding}) \\ 
        $r_{E,A}$ & 1500& [nM/day] & Expression rate of A when A enhancer is occupied (Eq.~\ref{ex1_F}) \\ 
        $r_{E,B}$ & 1000& [nM/day] & Expression rate of B when B enhancer is occupied (Eq.~\ref{ex1_F}) \\ 
        $r_{S,B}$ & 0& [nM/day] & Expression rate of B when B silencer is occupied (Eq.~\ref{ex1_F}) \\ 
        $r_{ES,B}$ & 500& [nM/day] & Expression rate of B when B enhancer and silencer are occupied (Eq.~\ref{ex1_F}) \\ 
        $k_m$ & 50& [day$^{-1}$] & Degradation rate of $m$=A, B (Eq.~\ref{ex1_F}) \\ 
        $\sigma(\z;\c)$ & $0.1 \begin{bmatrix}
            z_\text{A} & 0 \\ 0 & z_\text{B}
        \end{bmatrix}$ & [day$^{-1/2}$] & Noise magnitude of gene dynamics (Eq.~\ref{x_dynamics}) \\
        \hline
        \multicolumn{3}{l}{\textit{Example 2 Model Parameters}} \\ \hline
        $\beta_0$ & 1& [day$^{-1}$] & Division rate (Eq.~\ref{ex2_beta}) \\ 
        $\xi$ & 2.5& [$\mu$m] & Std. dev. of displacement of daughter from mother (below Eq.~\ref{ex2_beta}) \\ 
        $c_2^*$ & 23& [nM] & Threshold of $c_2$ above which cell division does not occur (Eq.~\ref{ex2_beta}) \\ 
        $\mu$ & 0.1& [day$^{-1}$] & Death rate \\ 
        $C_\ell$ & 1& [nM] & Parameter controlling excretion rate of signaling molecule $\ell=1,2$ \\ 
        $s_1$ & 7& [$\mu$m] & Std. dev. of spatial profile of excretion of signaling molecule 1 \\ 
        $s_2$ & 30 &[$\mu$m] & Std. dev. of spatial profile of excretion of signaling molecule 2 \\ 
        $r_0$ & 1000& [nM/day] & Expression rate of A when enhancer is unoccupied (Eq.~\ref{ex2_F}) \\ 
        $r_1$ & 2100 &[nM/day] & Enhancement in expression rate when enhancer is occupied (Eq.~\ref{ex2_F}) \\ 
        $K$ & 35& [nM] & Equilibrium const. for dimerization and heterodimer-Enhancer binding (Eq.~\ref{ex2_F}) \\ 
        $b$ & 250& [day$^{-1}$] & Degradation rate of protein A (Eq.~\ref{ex2_F}) \\ 
        $\sigma$ & $0.1z$& [day$^{-1/2}$] & Noise magnitude of gene dynamics (Eq.~\ref{x_dynamics}) \\ 
        $v_{0}$ & $10^3$& [$\mu$m$^2$/day] & Parameter controlling velocity of cell movement (Eq.~\ref{ex2_move}) \\ 
        $r_\text{min}$ & 8& [$\mu$m] & Spatial extent of attractive forces between cells (Eq.~\ref{ex2_move}) \\ 
        $\eta$ & 0& \:  & Noise magnitude of cell movement (Eq.~\ref{movement}) \\
    \end{tabular}
    \end{ruledtabular}
\end{table*}

\section{Summary and Conclusions}

We have systematically constructed a general model for tissue
development that incorporates intracellular gene expression dynamics,
the effects of gene expression state on proliferation and death rates,
heritability of cellular states during cell division, and cell-cell
interactions via chemical signaling. Dissecting our interacting-cell
framework into individual components, we provide a mathematically
unambiguous definition of a ``Waddington landscape'' (described as a
Waddington vector field to allow for nonequilibrium, driven dynamics)
and, if cell states are discretized, a developmental ``fitness
landscape.''

Our general framework integrates submodels for four processes:
stochastic changes in gene expression within each cell of the tissue
(Eq.~\ref{x_dynamics}), chemical signals produced and received by
cells affecting their gene expression dynamics
(Eq.~\ref{lambda_dynamics}), spatial movement of cells
(Eq.~\ref{movement}) which may affect their exposure to signaling
molecules, and finally, stochastic cell proliferation
(Eq.~\ref{reduced_master}) that allows daughter cells to acquire
different states than their mother (through the differential birth
rate parameter function of Eq.~\ref{birth}). These distinct processes
are coupled together through the chemical signaling field $\c$, the
drift and diffusivity of cells (which may depend on cell state $\z$
and chemical field $\c$), and the differential birth rate which may
also depend on mother-cell state and local chemical fields.  These
components represent an intermediate level of coarse-graining that
allows for better interpretability than fine-grained computational
models (such as CompuCell 3D \cite{compucell}) or unwieldy kinetic
theories \cite{Xia_JPA,Xia_PRE} that are difficult to marginalize to
obtain equations for interpretable and measurable quantities,
particularly if cells interact.

While the importance of cell-cell interactions and differential cell
division and death rates for robust development is well understood
\cite{discher2009growth,godard2019cell}, the complexity of
incorporating these features into mathematical models has been an
obstacle in modeling studies. Many theoretical works study cell fate
transitions for single cells \cite{wang2011quantifying,
  sisan2012predicting,fard2016not,rand2021geometry,
  schiebinger2021reconstructing,qiu2022mapping,
  frank2023macrophage,nakamura2024evolution,xue2024logic,chen2025quantifying}
or for populations of dividing but noninteracting cells without
spatial structure \cite{weinreb2018fundamental,fischer2019inferring,
  shi2019quantifying,zhang2021optimal,forrow2021lineageot,
  shi2022energy}. For example, cell-cell interactions mediated by a
chemical signaling concentration field, as in
Eq.~\ref{lambda_dynamics}, typically cannot be incorporated into
kinetic equations \cite{Xia_JPA,Xia_PRE,PhysicaD} in a straightforward
way.  Models that include spatial structure governed by cell-cell
interactions, but which do not incorporate tissue growth via
coordinated cell division or death, have also been studied
\cite{fooladi2019enhanced,smart2023emergent}. The models developed in
\cite{lei2020general,lei2023mathematical} incorporate cell-cell
interactions and differential cell division rates to describe
coordinated tissue development; these models also provide a more
detailed description of the cell cycle than our framework. However,
unlike our framework, they describe the cell population as a
continuous distribution of infinitely many cells rather than
describing individual cells as discrete entities, and they do not
incorporate spatial structure.

We applied our intermediate-grained stochastic dynamical framework to
two biological canonical processes: stem cell differentiation in a
well-mixed environment and a spatially-structured growth and
regeneration process.  While the examples in this work are simplified
models intended to provide conceptual clarity of general mechanisms of
robust tissue development, one should be able to apply our framework
to many other developmental processes that involve chemical signaling
such as morphogenesis, embryogenesis, organogenesis, and patterning
\cite{GILBERT,MORPH,MEIN}.

\appendix
\section{Mis-application of the Waddington Landscape Equations}

\begin{figure}
	\centering
	\includegraphics[width=\linewidth]{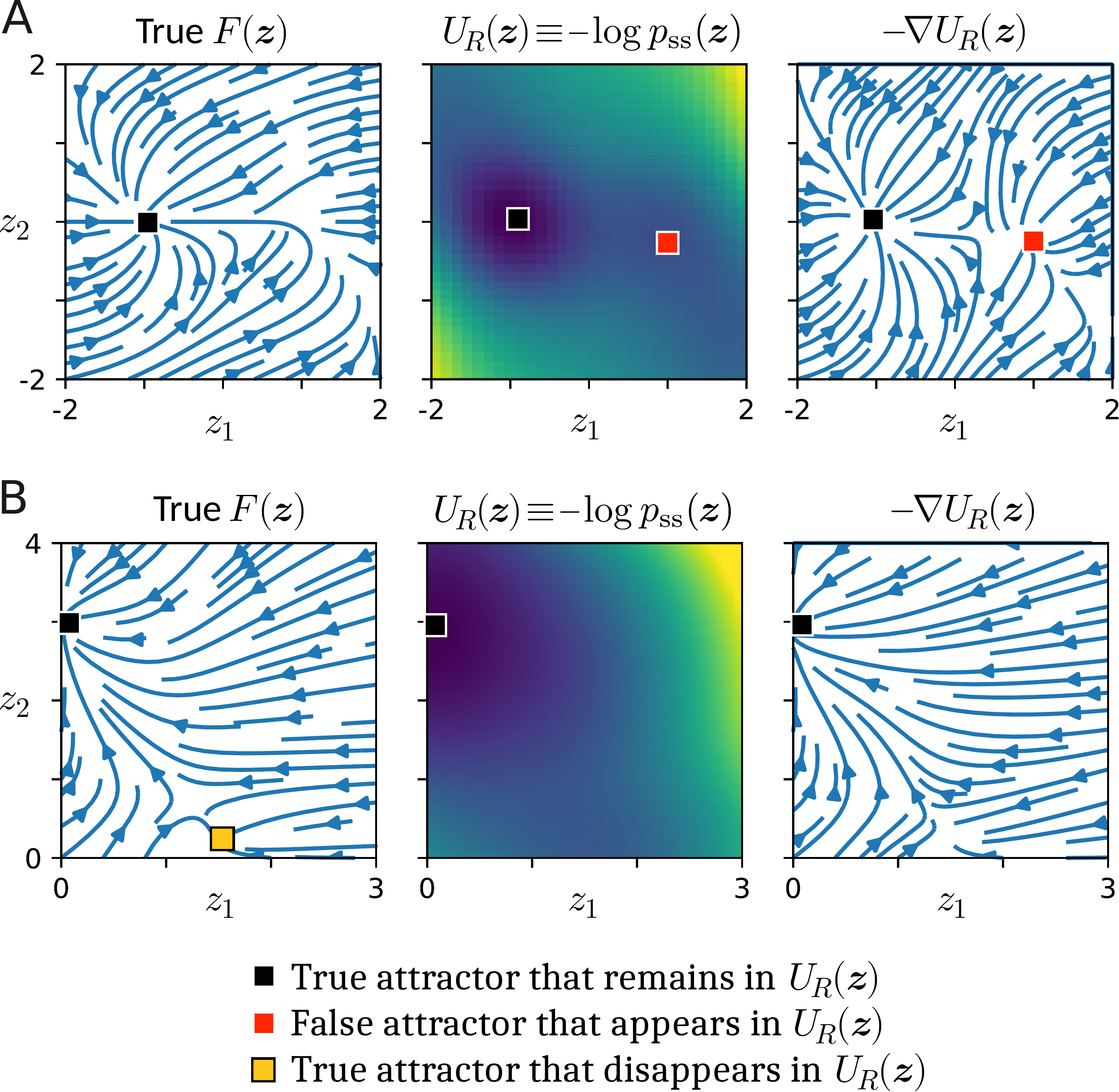}
	\caption{Artifacts due to the mis-application of
          Eqs.~\ref{grad_dynamics} and \ref{reconstructed}. (A) An
          example showing a false attractor (red square) which does
          not exist in the true vector field, but which appears in the
          reconstructed vector field. (B) An example showing a true
          attractor which disappears in the reconstructed vector field
          (orange square). The reconstructed landscapes $U_R(\z)$ are
          color coded where blue represents small values and yellow
          represents large values.}
	\label{U_R_fail}
\end{figure}

The mis-application of Eqs.~\ref{grad_dynamics} and
\ref{reconstructed} to systems that are either non-gradient systems,
or which have non-additive noise, can lead to artifacts in the
reconstructed dynamics. As in Eq.~\ref{reconstructed}, let
$U_R(\z)=-\log p_{\rm ss}(\z)$ denote a reconstructed landscape where
$p_{ss}(\z)$ is the steady state distribution of cell states generated
by some stochastic dynamics $\dd\z=\bm{F}(\z)\dd
t+\sigma(\z)\dd\W$. The reconstructed vector field $\bm F_R(\z)\equiv
-\nabla U_R(\z)$ can have `false' attractors (attractors that are
absent in $\bm{F}(\z)$. An example of this is shown in
Fig.~\ref{U_R_fail}A, with the false attractor indicated by a red
square. Additionally, $\bm F_R(\z)$ may fail to have attractors that
do exist in $\bm{F}(\z)$. An example of this is shown in
Fig.~\ref{U_R_fail}B, with the true attractor that disappears in the
reconstructed vector field indicated by an orange square.\\

\newpage

\bibliographystyle{unsrt}
\bibliography{main}
\end{document}